\documentstyle[12pt]{article}
\begin{document}
\title{Multidimensional Isotropic and Anisotropic q-Oscillator Models}

\author{ A. Ghosh   and P. Mitra\\
 Saha Institute of Nuclear Physics\\
 Block AF, Bidhannagar \\
Calcutta 700 064, India\\ \\A. Kundu\\
 Physikalisches Institut der Universit\"at Bonn\\
Nussallee 12, 53115 Bonn,
Germany\thanks{{\it Permanent address:
 Saha Institute of Nuclear Physics,
  AF/1 Bidhannagar,
Calcutta 700 064,India}}}

\date{}

\maketitle
\begin{abstract}
q-oscillator models are considered in two and higher  dimensions and
their  symmetries are explored. New symmetries are found for both
isotropic and anisotropic cases. Applications to the spectra of triatomic
molecules and superdeformed nuclei are discussed.
\end{abstract}
\newpage
\section{Introduction}
The  q-deformed  oscillator  introduced recently [1] has
spurred a great deal of activity. Various aspects of the
standard harmonic oscillator have been generalized to  the  q-deformed
case  [2].  In addition to studies on
models with a single q-oscillator and its successful  application
to different branches of physics [3-8],
higher-dimensional isotropic  and  anisotropic
q-oscillator  models too have begun to be studied [9-11].

It is known that the generalization of harmonic oscillator to q-oscillator
models becomes a nontrivial problem in higher dimensions, even in the
isotropic case. In [9] two such two-dimensional isotropic
q-oscillator models were proposed. However, only one of them
was really analysed,  a more natural candidate  being ignored.
 One of our aims
is to reconsider  this important issue and show that this alternative model
also exhibits an interesting symmetry. Our main
purpose however is to concentrate on the symmetries  of the anisotropic
q-oscillator models
in two and higher dimensions,
 which to our knowledge have not been explored properly.
It should be emphasized here that the symmetries of the {\it standard}
 (q=1) anisotropic oscillators in two dimensions are well understood [12].

We  derive  our  basic results first for $SU_q(N)$ with $N=2$
and subsequently extend them to higher  $N$. We find
 interesting applications of our results to
the  vibrational  spectra  of  triatomic  molecules  as  well  as
to the shell structure of superdeformed nuclei.

The plan of the paper is as follows. In Section 2 we review ordinary
oscillators, both isotropic and anisotropic, so that it will be
easier to make the transition to q-oscillators. In Section 3,
we discuss isotropic q-oscillators and in Section 4,
anisotropic q-oscillators. These two dimensional studies are
extended to three dimensions in Section 5.
Section 6 discusses some applications.
\section{Symmetries of standard oscillators}
Let  us begin by briefly discussing the
ordinary isotropic oscillator model in two dimensions.
The Hamiltonian
is given by
$$h=\sum_{i=1}^2\{a_i^\dagger,a_i\}=(n_1+n_2+1),\eqno(1)$$
where $n_i=a_i^\dagger a_i$ is the number operator and $a_i
^\dagger,~a_i~(i=1,2)$ are ladder operators with commutation relations
$$[a_i,a_j^{\dagger}]=\delta_{ij}, [a_i,a_j]=0, [n_i,a_i]=-a_i.\eqno(2)$$
"Angular momentum operators" may be constructed by the  Schwinger
representation
$$~j_+=a_1^{\dagger}a_2,~~ j_-=a_2^{\dagger}
a_1,~~ j_3={1\over 2}(n_1-n_2).\eqno (3)$$
They satisfy the  SU(2)  algebra and
commute  with  the  Hamiltonian  which is related to the Casimir
operator:  $$~j_{\alpha}j_{\alpha}={1 \over 4} (h^2-1).$$ This gives
the $SU(2)$
symmetry of the Hamiltonian (1).

Now  we ask what happens  when  the ordinary  oscillator
model (1) is made  anisotropic
giving  the Hamiltonian
$H~\propto~ \omega_1 (
 n_1 +{1\over2}) + \omega_2 (n_2 +{1\over2}).$
Of  course there is no degeneracy in general,
but the spectrum becomes interesting  when  the  frequencies
$\omega_1,\omega_2$ are
rational  multiples  of  each  other.  The  Hamiltonian  is  then
written as
$$H={1\over k_1}( n_1 +{1 \over2 })+{1\over k_2}( n_2 +{1 \over2 }).\eqno(4)$$
with $~\omega_1~:~\omega_2~=~{k_1}^{-1}~:~{k_2}^{-1}~$,where $k_1$ and  $k_2$
are
relatively  prime integers.
This  simple  situation   has been discussed in the literature
[12]. In this anisotropic  case  in  two  dimensions  an $ SU(2)$
 symmetry is involved
 again, though with a curious multiplicity $k_1k_2$ of  copies  of
each  irreducible  representation of the
corresponding algebra.
To explain the basic idea behind such symmetries we first note that the
 reason why the operators $~j_{\pm}$~  commute  with  the
Hamiltonian $~h $~ (1) is that they shift
both $~n_1,n_2$~ by unity but in opposite ways:
$$f(n_1,n_2)j_{\pm}=j_{\pm}f(n_1\pm 1,n_2\mp 1).$$
When we go on to the anisotropic Hamiltonian (4)
we need "angular momentum operators" which shift  $~n_1,  n_2$~  by
different amounts:
$$f(n_1,n_2) J_\pm= J_\pm f(n_1\pm k_1,n_2\mp k_2).\eqno(5)$$
Note that the single-quantum shift operator $e^{iP}$ which
may be associated with the  standard  creation  operator  through
$~a^\dagger  =\sqrt n e^{iP}$~
shifts $n$ by unity: $~~f(n)e^{iP}=e^{iP}f(n+1)$.
We  construct in analogy a multi-quanta shift
operator $~e^{ikP}$~  and  a  corresponding  multi-quanta  creation
operator   $~A^\dagger=\sqrt N e^{ikP}$~,   where   $~N=[{n \over k}]$~,   the
integral part of $~{n \over k}$~, i.e., $$~n= Nk+r, ~~~(0\leq r<k).$$
The operators $~A,  A^\dagger  $~  thus  constructed,  together with
their number
operator $~N$~, satisfy the usual oscillator algebra:
$$[A_i,A_j^{\dagger}]=\delta_{ij}, [A_i,A_j]=0, [N_i,A_i]=-A_i.\eqno(6)$$
Writing out  the  shift  operators  $~e^{ikP}=(n^{-{1\over 2}}a^\dagger)^k$~,
one finally arrives at
$$A^\dagger=\sqrt N
\left({(n-k)!\over n!}\right)^{{1\over 2}}a^{\dagger k}.\eqno(7)$$
Note  that  these  creation  and annihilation operators depend on
both $~N$~ and $~r$~ apart from the usual creation  and  annihilation
operators,  but  the  number  operator $~N$~ is independent of $~r$~.
These bosonic operators are the same as the  generalized
bosons  introduced  earlier in other contexts [12]. With their help
one may easily construct "angular momentum operators" [12] for  the
anisotropic  case  by  direct  analogy  with  the  isotropic  one:
$$ J_+=A_1^\dagger A_2,~~ J_-=A_2^\dagger  A_1, ~~ J_3={1\over
2}(N_1-N_2), \eqno(8)$$ where the appropriate $~k_i$~ are understood to
be used in the construction of $~A_i$~ from $~a_i~ (i=1,2)$~. It may be
noted that  unlike $~J_\pm$~, which depend on both $~N$~ and $~r$~ apart
from the operators $~a, a^\dagger $~, $~ J_3$~ depends only on $~N$~.

Due to the validity of the bosonic commutation relations (6), the
generators (8) provide again a Schwinger type realization of $SU(2)$  and
because of the modified shifts, they commute with the Hamiltonian (4).

The usual $SU(2)$ quantum numbers are given by
$~J={1\over 2}(N_1+N_2),   M={1 \over 2}(N_1-N_2)$~  and  the  energy  can  be
written as
$$E(J,r_1,r_2)=2J+{r_1 +{1\over2} \over k_1}+{r_2+
  {1\over2}\over k_2}.\eqno(9)$$
Thus the remainders $~r_1,r_2$~ enter the expression for the energy
here, which is not completely determined by $~J$~. These remainders
can take $~k_1$~ and $~k_2$~ different values respectively. Note  that
if  $~J~
,~M,~ r_1$~ and ~$r_2$~ are all specified then $~n_1, n_2$~
and hence the states are fixed. Now for given  values
of  $~r_1, r_2$~, the $ SU(2)$ quantum numbers $~J,M$~ can vary as usual
and every irreducible representation of the group occurs  exactly
once.  By  varying  $~r_1,  r_2$~,  one  therefore obtains $~k_1k_2$~
copies  of  each  irreducible  representation  of  $SU(2)$.   These
$~k_1k_2$~   copies   all have different energy, as is clear from
(9)
above.  Note  that
for  $~k_1=k_2=1$~ we recover the isotropic case with only one copy
of each irreducible representation.
Higher   dimensional  generalizations  can  be made by exploiting
these ideas of two dimensions and lead to  an  $SU(N)$  symmetry.
One interesting possibility should be noted here.  If  we  regard
the different levels for a fixed value of $~j$~ as forming a bunch,
the spread of energy values within a bunch may be so  large  that
the  bunches  overlap
violating the level sequence.  Whereas in the isotropic case levels with
higher values of $~j$~ have necessarily higher energies than levels
with  lower  $~j$~, in the anisotropic case, $~E(J,r_1,r_2)$~ exceeds
$~E(J+{1\over 2},r'_1,r'_2)$~ if $${(r_1-r'_1) \over k_1}+{(r_2-r'_2)\over
k_2} > 1.$$
This crossing of levels may be of interest in practical
situations.
\section{Isotropic q-oscillator models}
A q-oscillator involves
the q-commutators [ 1 ]
$$a^qa^{q\dagger}-qa^{q\dagger}a^q=q^{-n},~[n,a^q]=-a^q,\eqno(2')$$
where we take $q=e^{i\alpha}$  {\it not} to be a root of unity. One  can  use
two  independent  sets  of such operators to construct generators
$$j^q_+=a_1^{q\dagger}a_2^q,  j^q_-=a_2^{q\dagger}a_1^q,   j^q_3={1\over
2}(n_1-n_2)\eqno (3')$$ satisfying the $SU_q(2)$ algebra [ 1 ]
$$[j^q_+,j^q_-]=[2j^q_3]_q,~~[j^q_3,j^q_\pm]=\pm j^q_\pm.$$
Here $[x]_q$ stands for ${{q^x-q^{-x}}\over {q-q^{-1}}}={\sin{x\alpha}
\over  \sin{\alpha}}$.

A candidate [9] for the isotropic q-oscillator
Hamiltonian is (1) {\it itself} or its q-deformation: $$h^q_I=
[n_1+n_2+1]_q\eqno(10)
$$
All the generators of the $SU_q(2)$ algebra introduced above
commute  with  the Hamiltonian (10) and  in fact with an arbitrary
function of $n_1+n_2+1$. Thus, (1) itself, (10), or  an  arbitrary
function  of  $n_1+n_2+1$  and  $q$, which reduces to (1) when $q$
goes to unity, can be taken as the Hamiltonian for  the  isotropic
q-oscillator.  We  shall  refer  to such Hamiltonians as being of
Type  I.  Note  that  only the undeformed expression (or a linear
function thereof) can be expressed as a sum of two similar
terms for  the
two individual oscillators.
An interesting
property of such systems is that there  are  violations  of  the
level  sequence. This is due to
the  fact  that  an  inequality  $n'>n$  cannot   guarantee   the
inequality  $[n']_q>[n]_q$,  which holds only for small values of
$n, n'$ and $\alpha$. For such values,  the  energy  spectrum  is
similar  to  the  standard  case.  For  higher  values  of  these
quantities, however, the energy sequence is violated.
The dependence of $[n]_q={sin~n\alpha\over  sin~\alpha}$  on  the
parameter  $\alpha$ also gives a bound $\mid [n]_q\mid <\mid sin~
\alpha \mid^{-1}$, limiting the  energy  spectrum  in  a  bounded
range.  Different levels however will not coincide as long as $q$
is not a root of unity.

There  is  a more  natural  choice  for  the
Hamiltonian of the isotropic two-dimensional q-oscillator.
This   is  the  sum  of  the  Hamiltonians  for  two  independent
q-oscillators:
$$ h^q_{II}=    \sum_{i=1}^2\{a_i^{q\dagger},a_i^q\}=
[n_1+{1\over 2}]_q~+~[n_2+{1\over 2}]_q.\eqno(11)$$
As  mentioned above, this model
was briefly taken up in [ 4 ] where it was pointed out that
it does not commute with the generators  (3$'$) of the $SU_q(2)$
algebra. We shall refer to it as being of Type II. Note that its  spectrum
has obvious degeneracies arising from the permutibility of the
two q-oscillators.
 As we have restricted ourselves to
the situation where $q$ is not a root of unity, there  are
in general no multiplets besides  doublets (and singlets).
 It turns out that the permutation symmetry may induce
 an
interesting  $SU(2)$ or $SU_q(2)$
symmetry in this spectrum, as can be
seen in the following way. The algebra is made up of
the operators
\begin{eqnarray}\tilde    j_+ &=&\sum_{i=0}^\infty\sum_{n=1}^\infty
{[i]_q!\over
[i+n]_q!}(j_+^q)^n \mid i,i+n\rangle\langle i,i+n\mid \nonumber\\ &=&
\sum_{i=0}^\infty\sum_{n=1}^\infty\mid  i+n,i\rangle
\langle  i,i+n\mid,\nonumber\end{eqnarray}
$$\tilde  j_- =\tilde  j_+^\dagger,$$ $$
\tilde j_3 ={1\over 2}\sum_{i=0}^\infty\sum_{n=1}^\infty
[\mid i+n,i\rangle\langle i+n,i\mid - \mid i,i+n\rangle\langle
i,i+n\mid].\eqno(12)$$
Each term in the expansion of $\tilde j_+$ projects out the state
$\mid i,i+n\rangle$ and
in view of the representation (3$'$) and the relations
$$ a_i^q  \mid n_i\rangle = \sqrt{[n]_q} \mid n_i-1\rangle ,~
   a_i^{q\dagger}  \mid n_i\rangle = \sqrt{[n+1]_q} \mid n_i+1\rangle $$
$ (j_+^q)^n $ takes a state with $n_1=i, n_2=i+n$ to a state
with $n_1=i+n, n_2=i$ by creating $n$ quanta of the first kind
 while destroying an equal number of quanta of the second kind.
Similarly terms of $\tilde j_-$ act in the reverse way.
These operators satisfy the  $SU(2)$ algebra, which
can be  checked easily by direct calculation.
They even satisfy the $SU_q(2)$ algebra.
This happens because the operator $\tilde j_3$ has only zero  and
one-half as eigenvalues
as a result of the easily verifiable property $(\tilde j_+)^2=
(\tilde j_-)^2= 0$, and for these eigenvalues,
$SU(2)$  and  $SU_q(2)$  are  identical.
It can be checked easily that the above   operators
commute  with  (5) and therefore the  Hamiltonian can be
said to have an $SU(2)$  or  $SU_q(2)$  symmetry.
However,  this  symmetry  manifests  itself  only  in  doublets and
singlets, in contrast to the infinite variety of  representations
that is observed for the $SU(2)$ symmetry of (1) and the $~SU_q(2)~$
symmetry of (10).

\section{Anisotropic q-oscillators}

Let  us  go over to the anisotropic q-oscillator  of Type I.
The Hamiltonian $H_I^q$ in the isotropic  case  obviously  commutes
with   the   $SU_q(2)$   generators   introduced  above,  but  an
anisotropic       Hamiltonian
$$       H_I^q=[{1\over k_1} ( n_1+{1\over2})
+{1\over k_2} ( n_2+{1\over2}) ]_q,\eqno(13)$$
where $k_1,k_2$ are unequal,
positive integers having no common factor, does {\it not} do  so.
However one finds that,
much as before,  $q$- creation and annihilation operators
may be introduced through the unit quantum shift operator as
$a^{q\dagger}=\sqrt{[n]_q} e^{iP}$, and through the multi-quanta shift
operator as  $A^{q\dagger}=\sqrt{[N]_q} e^{ikP}$  giving
$$A_i^{q\dagger}=\sqrt{[N_i]_q}
\left(~{[n_i- k]_q!\over   [n_i]_q!}\right)^
{1 \over 2}~~(a_i^{q\dagger})^{k_i}.\eqno (14) $$
Here,  $[n]_q!=[n]_q[n-1]_q...[1]_q$  and  $N_i$  stands  for the
integral part  already  introduced  in  the  standard  case.  The
operators (14) proposed recently [13] as  generalized
$q$-bosons
can be used to construct generators
$$ J_+^q=
A_1^{q\dagger}A_2^q,{~~~}J_-^q=A_2^{q\dagger}A_1^q,{~~~}J_3^q={1\over
2}(N_1-N_2)\eqno(15)$$
satisfying the same $SU_q(2)$ algebra.
It is interesting to note that the generators $J_{\pm}^q,J_3^q$
thus constructed commute  with the Hamiltonian $H_I^q$.Therefore we
conclude that  anisotropic type I $q$-oscillator model (13)
exhibits again the $SU_q(2)$ symmetry as done by its isotropic
counterpart (10). Moreover,because
 of the splitting of  ${n_i\over  k_i}$
into  its  integral part $N_i$ and the fractional part ${r_i\over
k_i}$, exactly as before,  there  are  $k_1k_2$  copies  of  each
representation of $SU_q(2)$. These copies have different energies
as long as $q$ is  not  a  root  of  unity.  Thus,  although  the
nonlinear  expression  for the energy may change the ordering of the
levels
from the standard situation, the  degeneracies
remain exactly the same as before with the expression of energy
given by
$$E^q_I(J,r_1,r_2)=
[2J+{r_1 +{1\over2}\over k_1}+{r_2+ {1\over2}\over k_2}]_q.\eqno(9')$$

Next we come to the case of anisotropic q-oscillators of Type
II. The Hamiltonian is taken to be
$$ H^q_{II}=[{1\over k_1 } (n_1+{1\over 2})]_q+
[{1\over k_2}(n_2+{1\over 2}) ]_q.
\eqno(16)$$
By separating the integral and fractional parts as before, we can
write this as
$$ H^q_{II}= [N_1+{2r_1+1\over 2k_1}]_q+[N_2+{2r_2+1\over 2k_2}]_q.
\eqno(17)$$
However contrary to the isotropic case (11) this expression in general
does not exhibit the permutation
 symmetry under exchange of $N_1$ and  $N_2$,  because
of  the   fractions  which  are different in the two terms. It is
easy to see that these fractions can be equal only when
 $2r_i+1=k_i$ for $i=1,2$ i.e., only fixed values of $r_i$
 determined by  given $k_i$ can  lead to  degenerate levels.
  Thus there is a very
limited amount of degeneracy in the states of this system and  it
occurs  only if neither $k_i$ is even. When these restrictions on
$k_i$   and  $r_i$  are  satisfied,  there  occur  doublets  (and
singlets) much  as  in
the isotropic case of Type II.  There  is  again  an  $SU(2)$  or
equivalently  $SU_q(2)$ symmetry underlying these multiplets. The
corresponding
generators can be
constructed as in (12)
 with only the replacement of
the generators (3$'$) by the anisotropic ones (15) having
the properties
$$ A_i^q  \mid N_i\rangle = \sqrt{[N]_q} \mid N_i-1\rangle ,
   A_i^{q\dagger}  \mid N_i\rangle = \sqrt{[N+1]_q }\mid N_i+1\rangle $$
 Thus
 in terms of  the  operators  $J_{\pm}^q, J_3^q$, introduced
for  the  anisotropic  q-oscillator  of  Type  I  and appropriate
projection operators, we may write
\begin{eqnarray}\tilde    J_+ &=&
\sum_{I=0}^\infty\sum_{N=1}^\infty   {[I]_q!\over
[I+N]_q!}(J_+^q)^N\mid I,I+N;r_1,r_2\rangle\langle I,I+N;r_1, r_2\mid
\nonumber\\&=&
\sum_{I=0}^\infty\sum_{N=1}^\infty\mid  I+N,I;r_1,r_2\rangle
\langle  I,I+N;r_1,r_2\mid,
\nonumber\end{eqnarray}
 $$ \tilde  J_- =\tilde  J_+^\dagger,$$$$
\tilde J_3 ={1\over 2}\sum_{I=0}^\infty\sum_{N=1}^\infty
[\mid  I+N,I;r_1,r_2\rangle\langle   I+N,I;
r_1,r_2\mid
- \mid
I,I+N;r_1,r_2\rangle\langle
I,I+N;r_1,r_2\mid ].\eqno(18)$$
Here   the   states  ~are  understood  to  be  labelled  by  their
quantum numbers $N_1,~ N_2$,
$~ r_1,~r_2$
with fixed values $r_i={k_i-1\over 2}.$ Repeating similar arguments
it may be shown that
the  operators  introduced  in
(18)  not  only  obey  the  $SU(2)$ or $SU_q(2)$ algebra but also
commute with the Hamiltonian (16) and  thus  can  be  said  to  be
responsible  for  the degeneracy of the states. As in the case of
the corresponding isotropic q-oscillator, only the spin  one-half
(and  zero) representations occur instead of the infinite variety
observed for Type I.
\section{Higher dimensional models }
After this elaborate discussion of two dimensional oscillators,
its  generalization  to  higher  dimensions  does  not  pose  any
difficulty. For example, while the ordinary isotropic  three  dimensional
oscillator with the Hamiltonian $~H=n_1+n_2+n_3$~ exhibits an $SU(3)$
symmetry with certain irreducible representations (the  symmetric
tensor  ones) occurring exactly once, the anisotropic Hamiltonian
$~\tilde H={n_1 \over k_1}+{n_2 \over k_2}+{n_3 \over k_3}$~ leads to an
$ SU(3)$ with  each  of  the
above  representations  occurring $~k_1k_2k_3$~ times, the integers
$~k_i$~ being assumed to contain no overall common factor. Just  as
the  generators  of  the  algebra  in  the  isotropic case can be
written as $~a_1^\dagger a_2, a_2^\dagger a_3,  a_1^\dagger  a_3$~,
their conjugates and the elements of the Cartan subalgebra $~{1\over 2}(n_1-
n_2), {1\over 2}(n_2-n_3)$~, in the anisotropic case one replaces  $~a_i,
n_i$~ by $~A_i, N_i$~ defined with the appropriate $~k_i$~ involved in
the Hamiltonian. Repetition of  multiplets  occurs  because  the
energy  contains  a  new  piece  $~{r_1 \over k_1}+{r_2 \over k_2}+{r_3
\over k_3}$~, where
$~r_i$~ varies from 0 to $~k_i-1$~.

Apart from the possibility of overlapping bunches already seen in
the  case of two dimensions, a new pecularity may appear here. As
there are several $~k_i$~s, it may so happen that although there  is
no  overall common factor between them, some of them possess some
common   factor.   In  that  case,  there  will  exist  different
multiplets with equal energies. For instance, if  $~k_1=4,  k_2=2,
k_3=1$~,  the  copies  of  any  multiplet  with $~r_1=2, r_2=0$~ and
$~r_1=0, r_2=1$~  will  have  equal  energies.  Thus  the  symmetry
generators  do  {\it not  }  connect  all  states  having the same
energy.

More generally,  in  the  $\nu$   dimensional   case,   the   anisotropic
Hamiltonian  $\tilde H= \sum \limits_1^\nu\quad { n_i \over k_i}$
leads  to
$~\prod \limits_1^\nu k_i$~
copies of each irreducible representation of $ SU(N)$  observed  in
the  isotropic  case.  It  is  assumed  here  that the $~k_i$~s are
integers containing no overall common factor.  The  peculiarities
occurring in the lower dimensional cases can of course occur here
as well.

Similarly such higher dimensional cases can be considered with
q-deformations leading to an $SU_q(N)$ symmetry in both
the isotropic and anisotropic cases. The construction of symmetry operators
in the anisotropic case is analogous to the
undeformed situation; as in the two dimensional case,
the bosonic operators should be
replaced by their  q-deformations.

For the three dimensional q-oscillator of Type II
 one may take the generators as
$$\tilde j_{12}=\sum_{i=0}^\infty\sum_{n=1}^\infty
\sum_{n_3=0}^\infty\mid  i+n,i,n_3\rangle
\langle  i,i+n,n_3\mid, $$
$$\tilde j_{13}=\sum_{i=0}^\infty\sum_{n=1}^\infty
\sum_{n_2=0}^\infty\mid  i+n,n_2,i\rangle
\langle  i,n_2,i+n\mid, $$
$$\tilde j_3 ={1\over 2}\sum_{i=0}^\infty\sum_{n=1}^\infty
\sum_{n_3=0}^\infty
[\mid i+n,i,n_3\rangle\langle i+n,i,n_3\mid - \mid i,i+n,n_3\rangle\langle
i,i+n,n_3\mid],\eqno(19)$$
{\it etc} with the property $\tilde j_{ij}^2=0$.

\section{Some physical applications}
The three dimensional generalization of the Type I
model  has   interesting   physical   applications.   The
Hamiltonian in this case is
$$H_I^q=[{1\over k_1} ( n_1+{1\over2})
+{1\over k_2} ( n_2+{1\over2}) +{1\over k_3} ( n_3+{1\over2})
]_q. \eqno(20) $$
The generalized $SU_q(3)$ generators are given by
$$ J_{12}^q=
A_1^{q\dagger}A_2^q,{~~~}J_{13}^q=A_1^{q\dagger}A_3^q,{~~~}J_3^q={1\over
2}(N_1-N_2), \eqno(21) $$
{\it  etc}.  The  Hamiltonian  exhibits  $SU_q(3)$  symmetry with
a spectrum containing $k_1k_2k_3$ copies of each
symmetric tensor representation of $SU_q(3)$.
Considering $\alpha$ to be small, we expand (20) as
$$H_I^q
\approx (1+{\alpha^2\over 3!})H -{\alpha^2\over 3!}H^3,\eqno(22)$$
where $H$, the q=1 part of (20), is given by
$$\omega H=\omega_1 ( n_1+{1\over2})
+\omega_2 ( n_2+{1\over2}) +\omega_3 ( n_3+{1\over2})
\eqno(23) $$
with zero-order frequencies $\omega_i$ corresponding to its
normal modes. We find that there exists an
interesting connection between our model and the
spectra of triatomic molecules as well as  superdeformed nuclei.
Though there exist a number of investigations showing
the agreement of q-oscillator models  with  the  experimentally
observed  vibrational  and  rotational  spectra  of some diatomic
molecules with surprising  accuracy  [5,6],  similar  studies  on
triatomic  molecules  are  scarce, and even when available,  are
restricted to the isotropic case [10]. On  the  other  hand,  the
Hamiltonian (20) and hence (22) for small $q$-values can possibly
be  applied  to  explain  vibrational  spectra of a class of real
triatomic molecules with the  inclusion  of  anharmonicity  along
with  the  anisotropy   given in rational ratios. That is, for
triatomic molecules with normal modes in the ratios $$\omega_1  :
\omega_2   :  \omega_3=\epsilon_1  n_1:\epsilon_2  n_2:\epsilon_3
n_3,$$ where the $n_i$-s are integers  while  $\epsilon_i$-s  are
numbers  close  to unity showing deviations from rational ratios.
For example, our model is expected to  describe  the  vibrational
spectra of molecules like [15]
\begin{enumerate}
\item HOCl, with zero-order frequencies (in cm$^{-1}$)
$$\omega_1=3609,~~ \omega_2=1238, ~~\omega_3=720,\eqno(24a)$$
in the ratio
$\omega_1   :   \omega_2   :  \omega_3\approx 5:2:1$
with accuracy
$$\epsilon_1 =1.00,~~\epsilon_2 =0.86, ~~\epsilon_3=1.00.\eqno(24b)$$

\item HDO, with zero-order frequencies (in cm$^{-1}$)
$$\omega_1=2724,~~ \omega_2=1403, ~~\omega_3=3707,\eqno(25a)$$
in the ratio
$\omega_1   :   \omega_2   :  \omega_3\approx 2:1:3$
with accuracy
$$\epsilon_1 =0.97,~~\epsilon_2 =1.00, ~~\epsilon_3=0.88.\eqno(25b)$$

\item H$_2$O, with zero-order frequencies (in cm$^{-1}$)
$$           \omega_1=3825.3,\quad          \omega_2=1653.9,\quad
\omega_3=3935\eqno(26a)$$
in the ratio
$\omega_1   :   \omega_2   :  \omega_3\approx 2:1:2$
with accuracy
$$\epsilon_1 =1.16,\quad \epsilon_2 =1.00,\quad \epsilon_3 = 1.19.\quad
\eqno(26b)$$
\end{enumerate}
For demonstrating our claim, we present here in detail the case
of the water molecule.
The vibrational spectra of H$_2$O molecules (without considering
the anharmonicity) may be
described by the Hamiltonian (23) [14,15] rewritten as
$$\omega H=\omega\left(\epsilon_1( n_1+{1\over2})
+{\epsilon_2 \over 2} ( n_2+{1\over2}) +\epsilon_3 ( n_3+{1\over2})\right)
\eqno(27) $$
with
$ \omega=3307.8$ and $\epsilon_i$ as given in (26b).
We see that  our anisotropic
model (20) for $k_1:k_2:k_3= 1:2:1$ and $q=1$ can describe this
system with fairly good accuracy. The system thus has an approximate
$SU(3)$ symmetry when anharmonic effects are neglected.

For describing the experimental result more accurately
anharmonic terms are usually considered and for triatomic molecules like
H$_2$O the number of such anharmonic parameters is six. We show
 in Table 1 that
 our single-parameter q-oscillator model  can also  describe
such anharmonic effects with good accuracy.
The energy spectrum obtained
from our formula (22) including anharmonicity
is shown in the penultimate column
of Table 1. Note its close
resemblance with the spectrum of H$_2$O given in the last
column, which is
calculated from the 6-parameter fit of [15].
Here  E(anh-q) is a {\it single} parameter fit,
where $\alpha$=0.0849 is chosen to get  the best fit
with the 6-parameter result within the given range.

This suggests that the vibrational Hamiltonian of H$_2$O
molecules with anharmonicity
exhibits an approximate $SU_q(3)$ symmetry (with $q=\exp^{i 0.0849}$)
 along with all the interesting features of the anisotropic
q-oscillator model discussed here.

As another possible application, one should mention that in analysing
the shell structure of superdeformed nuclei one usually considers the
energy spectrum [16]
$$E(n_1,n_2,n_3)=\hbar\omega_{\perp} ( n_1+
 n_2+1) +\hbar\omega_3 ( n_3+{1\over2})
\eqno(28). $$
The major shell structure is observed only when $\omega_{\perp}$
and $\omega_3$ are in the ratios of small  integers.  The  reason
behind   this  fact  and  the  symmetry  involved  are  not  well
understood, as stressed by Mottelson [16]. We see here that such
superdeformed nuclear models with nonlinear terms may
 well be represented by our
anisotropic q-oscillator with an $SU_q(3)$ symmetry, which may be
a preferred symmetry providing a stable structure.

\bigskip \section*{Acknowledgements}
The basic problem was suggested by Prof. Avinash Khare. A.G. would like
to  thank  him  for  many useful discussions.
One of the authors (AK) thanks Alexander von Humboldt-Stiftung for financial
support.

\newpage

\begin{center}
\begin{tabular}{|r|r|r|p{.5in}|p{.5in}|p{.5in}|}
\hline
\multicolumn{1}{|c|}{ {\sl n1} }& \multicolumn{1}{c|}{ {\sl n2} }
& \multicolumn{1}{c|}{ {\sl n3} }   & \multicolumn{1}{c|}{{ \sl E0 }}
& \multicolumn{1}{c|}{{ \sl E(anh-q)}}
& \multicolumn{1}{c|}{{ \sl E(exp)}}
\\
\hline
    0&   0 &  0&   4707 &  4701 &  4680\\
    0   &1&   0&   6361&   6340 &  6295\\
    0&   2&   0&   8015&   7968 &  7871\\
    1 &  0&   0&   8531 &  8473 &  8419\\
    0&   0&   1&   8643&   8583 &  8524\\
    0&   3 &  0&   9669&   9581 &  9408\\
    1&   1&   0&  10185&  10081 & 10102\\
    0&   1&   1&  10297&  10190 & 10121  \\
    0&   4&   0&  11323&  11177 & 10906\\
    1&   2&   0&  11839&  11671 & 11746\\
    0&   2&   1&  11951&  11778 & 11678 \\
    2  & 0&   0&  12355 & 12162 & 12075\\
    1&   0&   1&  12467&  12269 & 12107\\
    0&   0&   2&  12580&  12376 & 12277\\
    0 &  5 &  0 & 12977  & 12752 & 12365 \\
    1 &  3 &  0 & 13493 & 13239 & 13351 \\
    0 &  3 &  1 & 13605 & 13345 & 13197\\
    2 &  1 &  0 & 14009  & 13724 & 13825\\
    1 &  1 &  1 & 14121  & 13829 & 13771\\
    0 &  1 &  2 & 14233 & 13934 & 13855\\
    1 &  4 &  0 & 15146 & 14783 & 14917 \\
    0 &  4 &  1 & 15259 & 14887 & 14676\\
    2 &  2 &  0 & 15662 & 15259 & 15537\\
    1 &  2 &  1 & 15775 & 15363 & 15396\\
    0 &  2 &  2 & 15887 & 15466 & 15393\\
    3 &  0 &  0 & 16178 & 15733  & 15645\\
    2 &  0 &  1 & 16291  & 15836  & 15605 \\
    1 &  0 &  2 & 16403 & 15938 & 15702\\
    0 &  0 &  3 & 16516 & 16041 & 15937\\
\hline
\end{tabular}
\end{center}
\smallskip
\begin{itemize}
\item [Table 1:]  {\it Energy spectrum $E= \omega H^q_I$, where $H^q_I$
is as in (22) with $\alpha=0.0849$.
$E(anh-q)$ and  $E0$   give the values
with and  without   consideration of the anharmonicity,
while $E(exp)$ was obtained from [15].}
\end{itemize}

\newpage
\centerline {\bf REFERENCES}\bigskip
\begin{itemize}
\item [1.]
A. J. Macfarlane, J. Phys. {\bf A22}  (1989)  4581;
L. C. Biedenharn, J. Phys. {\bf A22}  (1989)  L873
\item [2.] C. P. Sun and H. C. Fu, J. Phys. {\bf A22}  (1989)  L983;
T. Hayashi, Comm. Math. Phys. {\bf 127}  (1990)  129;
Y. J. Ng, J.Phys.{\bf A 23}  (1989)  1023;
\item [3.] M. Chaichian, D. Elinaas and P. Kulish, Phys. Rev. Lett.
{\bf 65}  (1990)  980
\item [4.] R. W. Gray and C. A. Nelson, J.Phys. {\bf A23}  (1990)  L945;
\item [5.] Z. Chang and H. Yan, Phys. Lett. {\bf A 158}  (1991)  242
 \item [6.] D.Bonatsos, P.Raychev and  R.Roussev
    Chem. Phys. Lett. {\bf 175} (1990) 300
 \item [7.] P.Raychev, R.Roussev and Yu.F.Smirnov
 J.Phys. {\bf G16} (1990) L137
 \item [8.] M.Kibler et al, J. Phys. {\bf 24A} (1991) 5283
\item [9.]
E. G. Floratos, J. Phys.  {\bf A 24}  (1991)  4739;
\item [10.] A. Kundu and Y. J. Ng, Phys. Lett. {\bf A197} (1995) 221
\item [11.] D. Bonatsos, C. Daskaloyannis, P. Kolokotronis
and D. Lenis, {\it Oscillator with rational ratios of
frequencies and the Nilsson model}, report no. hep-th/9411218 (Nov. 1994)
\item [12.] R. A. Brandt and O. W. Greenberg, J. Math. Phys. {\bf 10}  (1969)
1168
A. Cisnerius and H. Macintosh, J. Math. Phys.{\bf 11}  (1970)  870
\item [13.] J. Katriel and A. I. Solomon,J. Phys. {\bf A24}  (1991)  2093
\item [14.] G. Herzberg, Molecular Spectra and Molecular Structure
(D. Van Nostrand, 1951)
\item [15.]
M. A. H. Smith, C. P. Rinsland, B. Fridovich and K. N. Rao,
{\it Molecular Spectroscopy}: Modern Research  vol III
(ed. K.N. Rao, Acad. Press, 1985) p. 155
\item [16.] B. Mottelson, Nucl. Phys. {\bf A522} (1991) 1c-12c
\end{itemize}
\end{document}